\documentstyle[twoside,fleqn,espcrc2,epsf]{article}

\newcommand{\AmS}{{\protect\the\textfont2
  A\kern-.1667em\lower.5ex\hbox{M}\kern-.125emS}}

\title{The Structure of Flux Tubes in Maximal Abelian Gauge}

\author{Christoph Schlichter\address{Fachbereich Physik, Bergische
        Universit\"at, D-42097 Wuppertal, Germany}%
        \thanks{Talk presented by Christoph Schlichter.},
        Gunnar S.\ Bali\address{Physics Department, The University,
        Highfield, Southampton SO17 1BJ, United Kingdom},
and
         Klaus Schilling\address{HLRZ c/o Forschungszentrum J\"ulich,
        D-52425 J\"ulich and DESY, 22603, Hamburg, Germany}}
       
\begin{document}

\begin{abstract}
Various field strength correlators are investigated in the maximal
abelian projection of pure SU(2) lattice gauge theory. High precision
measurements of the colour fields, monopole currents, their curl and
divergence allow for detailed checks of the dual superconductor
scenario. A further decomposition of abelian observables into
monopole and photon parts reveals that the flux tube is built up from
the monopole part alone.
\end{abstract}

\maketitle

\section{INTRODUCTION}
\noindent
Almost 20 years ago, 't Hooft and Mandelstam proposed the dual
superconductor scenario of confinement: it is believed that at low
temperature magnetic monopoles condense and thus, chromo electric flux
is expelled from the QCD vacuum, analogous to a type II
superconductor.  This leads to the formation of thin electric flux
tubes between colour charges and thus to a linearly rising potential,
explaining confinement. In view of both, making genuinely non
perturbative aspects of QCD accessible to analytical treatment and
understanding how confinement arises, it is worthwhile to investigate
to what extent QCD reproduces expectations from the Ginzburg-Landau
(GL) equations, and to fix their relevant parameters.  Pioneering
steps in this direction have been made in Refs.~\cite{haymaker,cea}.

In nonabelian theories, the definition of fields and currents becomes
ambiguous. 't Hooft suggested the maximal abelian gauge as a
renormalizable gauge condition for projecting out a Cartan subgroup,
believed to be relevant for infra red aspects of QCD (abelian
dominance).  In this investigation, we follow these lines.
Previously, the abelian string tension has been found to reproduce the
nonabelian one within 8\%~\cite{shiba,bali_vborn_mmp_ks}. A further
decomposition of the abelian Wilson loop into photon and monopole
parts~\cite{smitvds} shows that the monopole part accounts exclusively
for the string tension~\cite{leewoltrot,bali_vborn_mmp_ks}.  Here we
extend these studies to the level of field distributions.

\section{SIMULATION}
\noindent We investigate a $32^4$ lattice at $\beta=2.5115$,
corresponding to a lattice spacing $a\approx 0.086$~fm, where the
scale has been set from the string tension value $\sqrt{\kappa}=440$~MeV.
A careful study of systematic errors introduced by
incomplete gauge fixing and optimisation of the gauge fixing
procedure are crucial for obtaining
reliable results. We take measurements on 108 independent
abelian configurations, generated in the context of
Ref.~\cite{bali_vborn_mmp_ks}. 
In the limit of large $T$, we measure
connected correlation functions~\cite{prd51colourflux},
\begin{equation}\label{Flusskorrelator}
\langle O ({\mathbf x},T) \rangle = \frac{\langle O ({\mathbf
x},T/2)W(R,T)\rangle}{\langle W(R,T) \rangle}-\langle O\rangle.
\end{equation}
By varying ${\mathbf x}$, we scan through the space around
two static sources, separated by a
distance $Ra$.
Depending on the type of the local probe $O$, distributions of
${\mathbf E}$, ${\mathbf B}$, $\nabla^+\wedge {\mathbf E}$,
$\nabla^-\wedge\nabla^+\wedge {\mathbf E}$, ${\mathbf J}_m$ and
$\nabla^-\wedge {\mathbf J}_m$ are obtained.
$W$ denotes a smeared Wilson loop, which is either built up
from abelian link angles or from the regular (photon) or the singular
(monopole) contributions to these angles alone, according to the
approximate factorisation of Ref.~\cite{bali_vborn_mmp_ks}.

\section{MONOPOLE CONTRIBUTION}
\begin{figure}[h]
\vspace{-1cm}
\epsfxsize=8cm\epsfbox{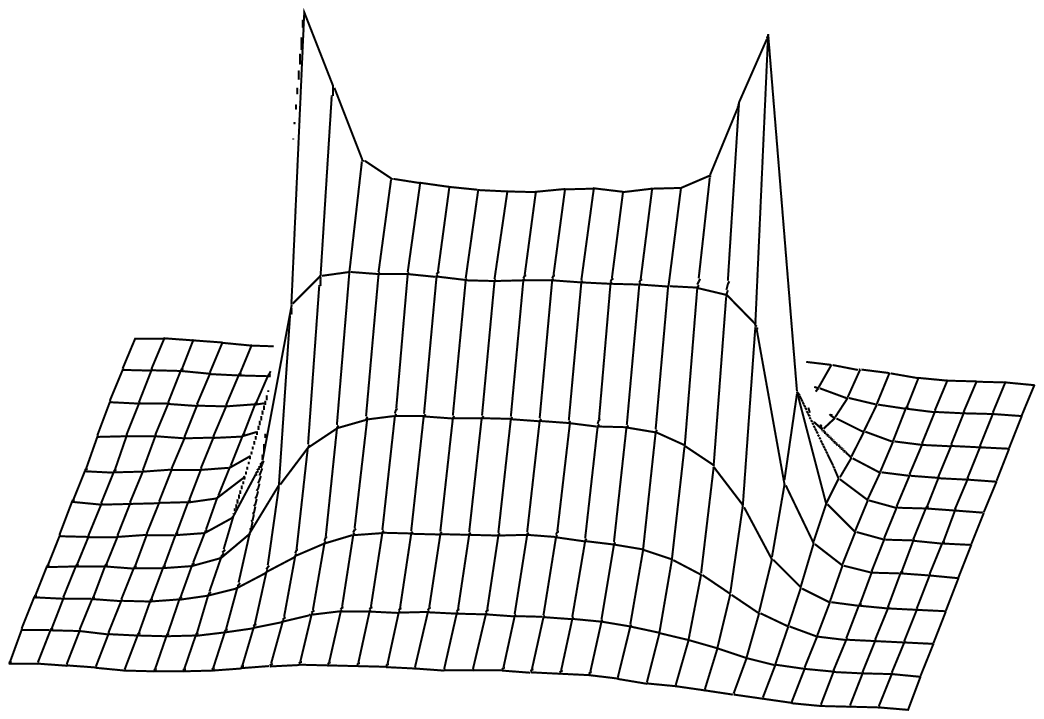}
\vspace{-2cm}
\epsfxsize=8cm\epsfbox{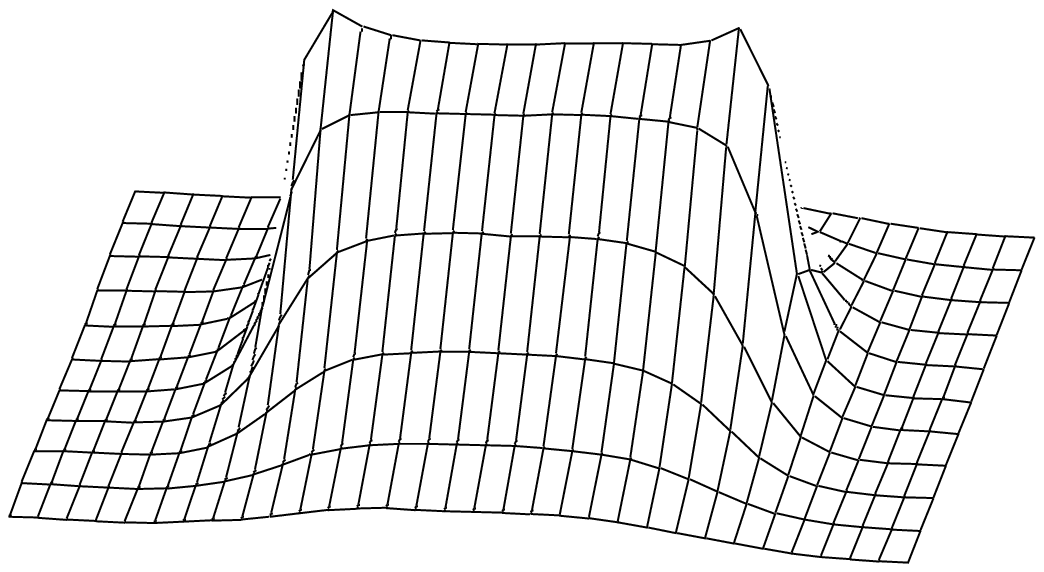}
\vspace{-1.8cm}
\epsfxsize=8cm\epsfbox{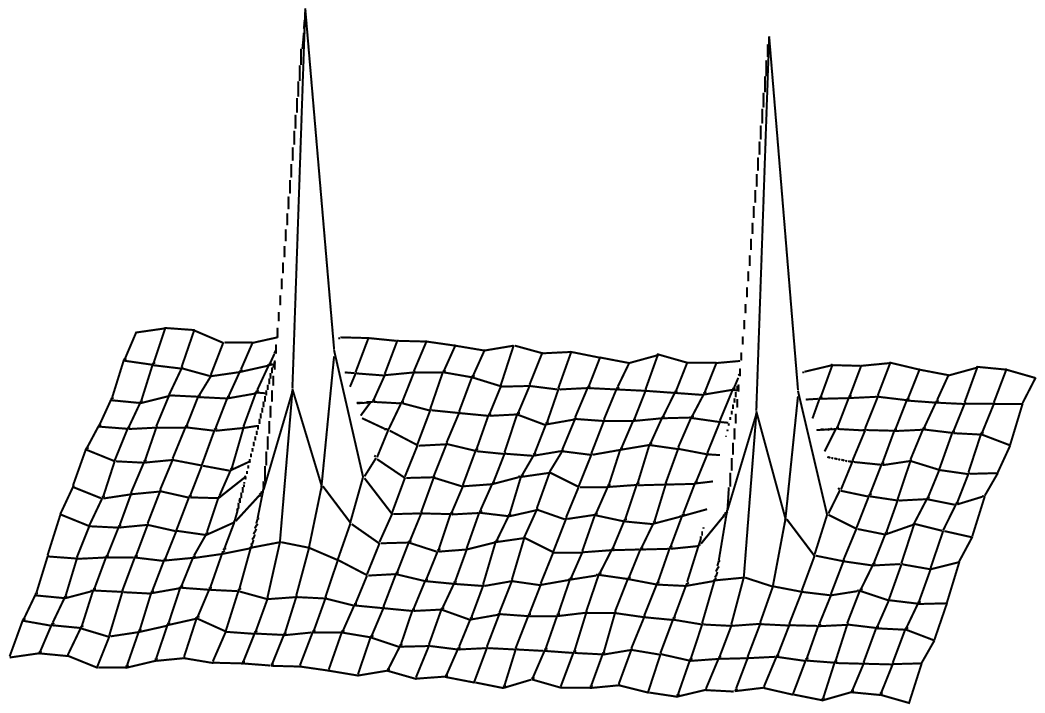}
\vspace{-1.8cm}
\caption{Abelian action density: full distribution,
monopole and photon contributions, respectively.\vskip -0.8cm}
\label{densities}
\vspace{3ex}
\end{figure}

\noindent
It has been demonstrated previously that the abelian potential
approximately factorises into a monopole and a photon
contribution~\cite{leewoltrot,bali_vborn_mmp_ks}.  Here, we
investigate this factorisation on the level of the underlying field
distributions.  In Fig.~\ref{densities} the situation is depicted for
the action density, $\sigma=1/2({\mathbf E}^2-{\mathbf B}^2)$.  In
accord with na\"\i{}ve expectations, the string is entirely due to the
monopole part, whereas the effect of the photon part is localised in
the vicinity of the sources\footnote{A database of colour images can be
accessed via anonymous ftp from wpts0.physik.uni-wuppertal.de. The
compressed .rgb and .ps files are deposited in the directory
/pub/MAcolorflux.}.

\section{DUAL SUPERCONDUCTIVITY}
\noindent
We choose the coordinates such that the colour sources
lie on the $z$-axis at positions $z=\pm r/2$
with $r=Ra$.
$x_{\perp}=(x^2+y^2)^{1/2}$ denotes the
transverse distance from the interquark string.
A combination of Ampere's law $\nabla\wedge{\mathbf E}={\mathbf J}_m$
with the London
equation ${\mathbf A}+\lambda^2{\mathbf J}_m=0$ yields
$E_z(x_\perp)-\lambda^2\Delta_\perp E_z(x_\perp)=\Phi_m\delta^2(x_\perp)$.
We expect the analytical solution,
\begin{equation}\label{bessel}
E_z = \frac{\Phi_{m}}{2 \pi \lambda^2} K_{0}(x_{\perp}/\lambda),
\end{equation}
to hold in the London limit which should be realized at sufficiently
large $r$ and $x_{\perp}$. For $T\geq 6$, we find all our data
on $\nabla^+\wedge{\mathbf E}$ and ${\mathbf J}_m$ to be
in perfect agreement with
Ampere's law and proceed to fit $E_z$ in the central plane ($z=0$)
to Eq.~(\ref{bessel}) as a test of the validity of the London equation.

\begin{table}[hbt]
\vskip -.6cm
\setlength{\tabcolsep}{0.8pc}
\catcode`?=\active \def?{\kern\digitwidth}
\caption{Fits of $E_z$ at $R=8$ according to Eq.~(\protect\ref{bessel}).}
\label{tab1}
\begin{tabular}{lrrr}
\hline
range & $\Phi_m$ & $\lambda$ & $\chi^2/{\mbox{dof}}$ \\ \hline
1\ldots 7.07 & 2.2(3) & 3.6(5) &  1700 \\
2\ldots 7.07 & 1.7(1) & 2.4(1) &  150 \\
3\ldots 7.07 & 1.7(1) & 2.0(1) &  6 \\
3.1\ldots 7.07 & 1.8(1) & 2.0(1) &  9 \\
3.6\ldots 7.07 & 1.9(1) & 1.9(1) & 4 \\
4\ldots 7.07 & 2.0(1) & 1.88(6) &  1.8 \\
4.2\ldots 7.07 & 1.9(1) & 1.82(7) &  1.2 \\
4.5\ldots 7.07 & 2.0(1) & 1.82(2) &  1.1 \\
5\ldots 7.07 & 2.1(1) & 1.66(7) &  0.7 \\
\hline
\end{tabular}
\vskip -.7cm
\end{table}

\begin{figure}[htp]
\vskip -.7cm
\unitlength=1cm
\begin{picture}(8,6)
\put(0,0.2){\mbox{\epsfxsize=7.5cm\epsfbox{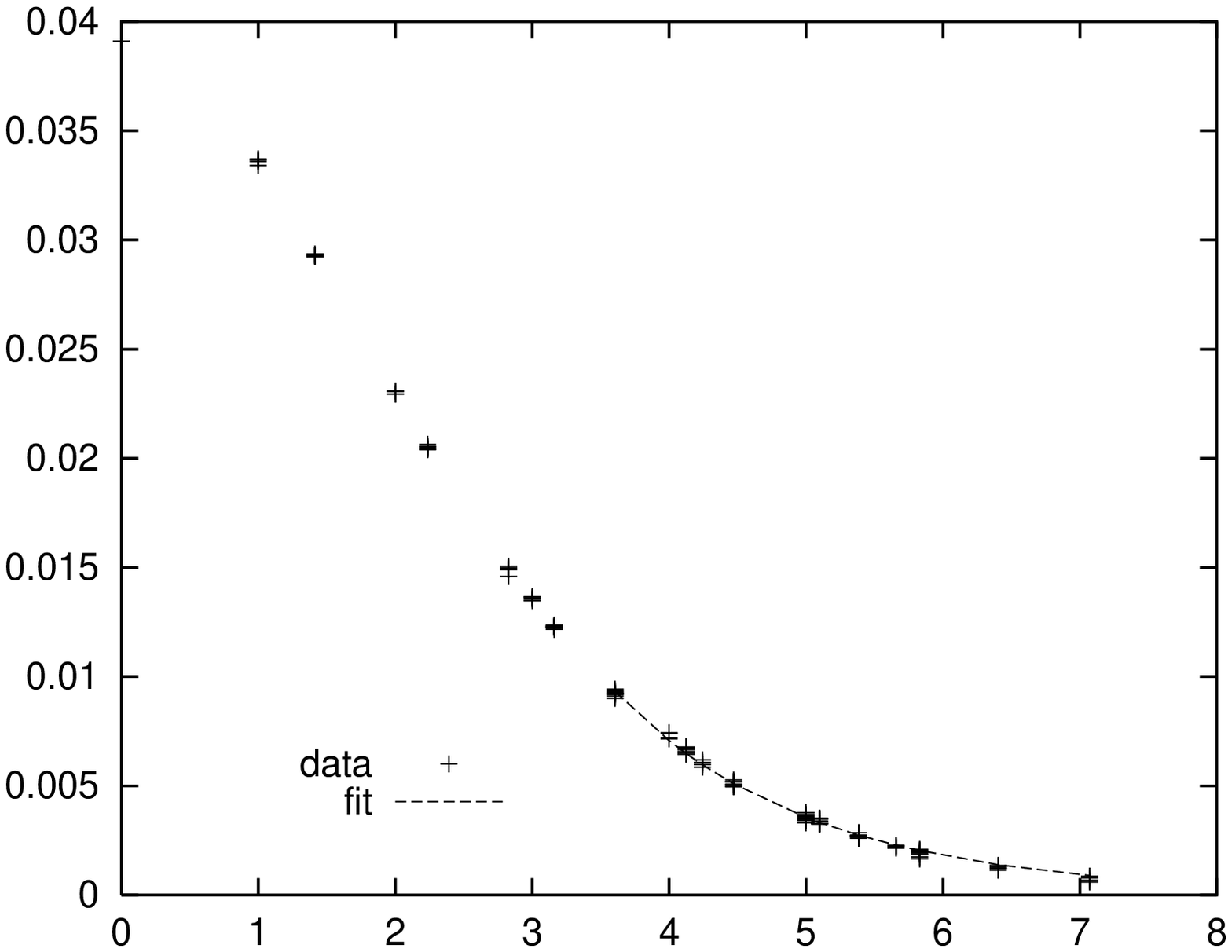}}}
\put(3.5,2.7){\mbox{\epsfxsize=3.5cm\epsfbox{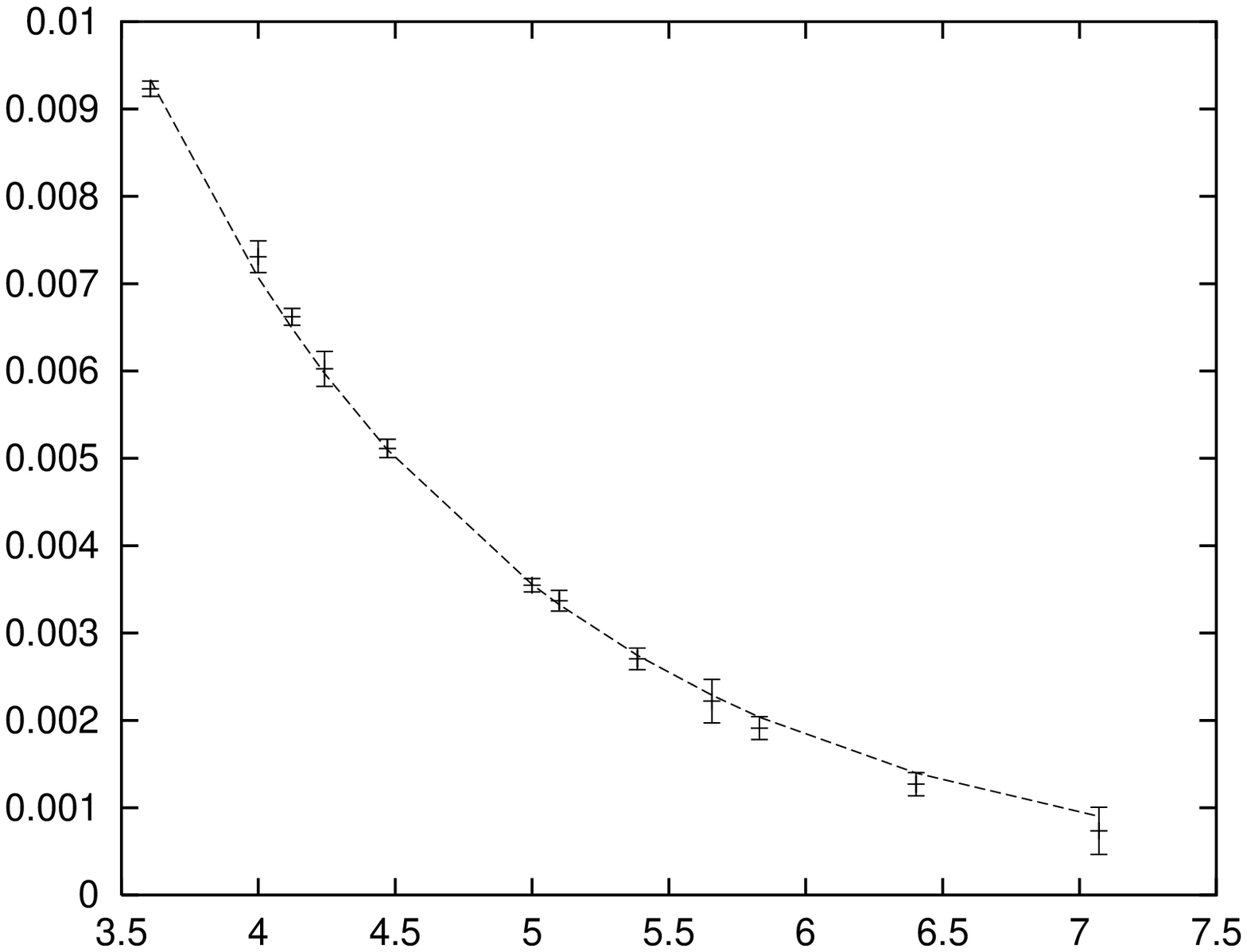}}}
\put(6.6,0){\mbox{$x_\perp$}}
\put(0,5){\mbox{$E_z$}}
\end{picture}
\vskip -.8cm
\caption{Fit of $E_z$ to Eq.~(\protect\ref{bessel})
(for $R=10$).\vskip -.6cm}\label{fig2}
\end{figure}

Results for various fit ranges
are
summarised in Table~\ref{tab1}
for the example $R=8$.
For $x_{\perp}\ge0.3$~fm, the fit parameters
become stable and the $\chi^2$-values reasonable. 
One such fit (for $R=10$) is displayed in Fig.~\ref{fig2}.
For all distances, we obtain penetration lengths of $\lambda\approx 0.15$~fm.
The value of the external magnetic flux $\Phi_m$ is determined by the
magnetic charge of
the objects, into which the monopoles condense.
Assuming the London limit to apply we find $\Phi_m\approx 2$.

We proceed to compare the small $x_{\perp}$ data directly to the
expectation from dual GL equations that relate the vector potential
${\mathbf A}$, to the magnetic monopole current ${\mathbf J}_m$. The
GL wave function $\psi=\psi_{\infty}f(x_{\perp})e^{i\theta}$ and its
coherence length $\xi$ enter the scenario.  $f$ approaches zero as
$x_{\perp}\rightarrow 0$.  The London limit corresponds to
$f(x_{\perp})=1$. This is the asymptotic value as
$x_{\perp}\rightarrow\infty$.  The ansatz $f(x_{\perp})=\tanh(\nu
r/\xi)$ approximately solves the nonlinear GL equations for these
boundary conditions.  A small-$x_{\perp}$ expansion yields $\nu
=\sqrt{3/8}$. We determine the vector potential by
numerically integrating the electric field,
\begin{equation}
A_{\theta}(x_{\perp})
=\frac{1}{x_{\perp}}\int_0^{x_{\perp}}\!dx'_{\perp}\,x'_\perp
E_z(x'_\perp ). 
\end{equation}
Subsequently, this data is fitted to,
\begin{equation}
A_{\theta}(x_\perp)=\frac{\lambda^2}{f(x_\perp)^2}
J_{m,\theta}(x_\perp)-\frac{\Phi_m}{2\pi x_{\perp}},\label{gl}
\end{equation}
with parameters $\Phi_m$, $\xi$ and $\lambda$. $J_{m,\theta}$ is a
parametrisation fitted to the
angular component of the measured magnetic monopole current.

\begin{table}[hbt]
\vskip -.6cm
\setlength{\tabcolsep}{0.8pc}
\catcode`?=\active \def?{\kern\digitwidth}
\caption{Fits of $A_{\theta}$ and $J_{m,\theta}$ to Eq.~(\protect\ref{gl})
(at $R=8$).}
\label{tab2}
\begin{tabular}{lrrr}
\hline
$T$ & $\lambda$ & $\xi$ & $\Phi_m$ \\ \hline
2 & 1.88(4) & 1.19(2) & 1.03(1) \\
3 & 1.88(3) & 1.56(2) & 1.20(1) \\
4 & 1.84(2) & 1.76(2) & 1.28(1) \\
5 & 1.75(2) & 2.06(2) & 1.35(1) \\
6 & 1.82(2) & 2.05(2) & 1.41(1) \\
\hline
\end{tabular}
\vskip -.7cm
\end{table}

\begin{figure}[htp]
\vspace{5.5cm}
\begin{picture}(8,6)
\unitlength1cm
\put(0,0){\mbox{\epsfxsize=7.5cm\epsfbox{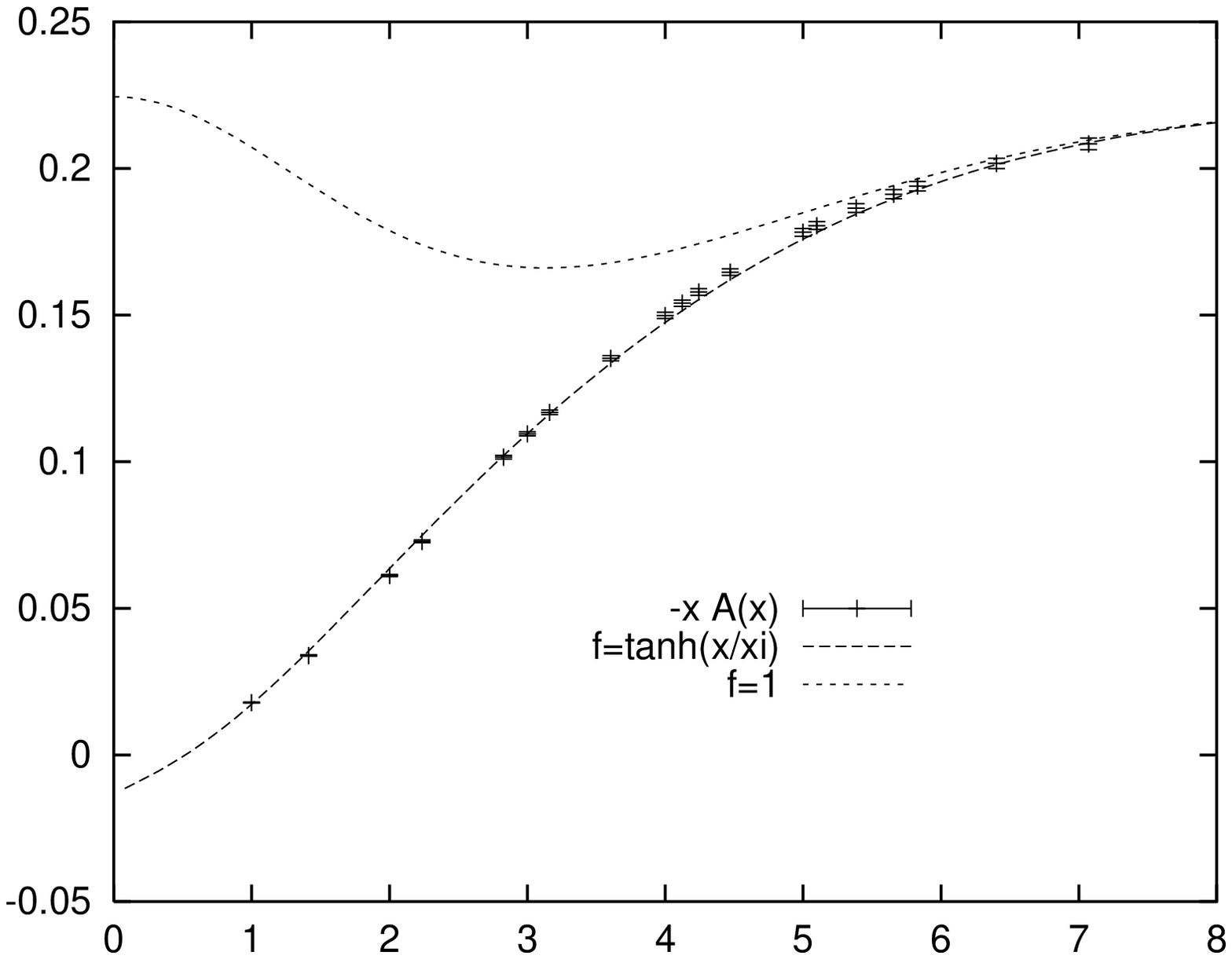}}}
\put(6.6,-0.2){\mbox{$x_\perp$}}
\put(-0.2,4.7){\mbox{$-x_\perp A_\theta$}}
\end{picture}
\vskip -.8cm
\caption{Check of Eq.~(\protect\ref{gl}) for $R=8$, $T=6$.
The $f(x)=1$ curve corresponds to the London limit.\vskip -.6cm}\label{fig3}
\end{figure}

All data are well described by the fits.  Results for the example
$R=8$, corresponding to a source separation $r\approx 0.7$~fm, are
summarised in Table~\ref{tab2}.  The quality of the fit for $T=6$ is
visualised in Fig.~\ref{fig3}.  The difference between the $f=1$ curve
and the data indicates to what extent we are probing surface effects.
The results on the penetration length nicely agree with results from
Eq.~(\ref{bessel}). A plateau in $\xi$ is reached for $T\geq
5$. However, the value for $\Phi_m$ is not saturated yet; we obtain a
lower limit, $\Phi_m\le1.4$. The ratio, $\lambda/\xi=0.89(2)$ turns out
to be definitely larger than the critical value $1/\sqrt{2}$, i.e.\
the SU(2) appears to act like a type II superconductor~\cite{diss}.

\section*{ACKNOWLEDGEMENTS}
\noindent
We acknowledge support by the DFG (grants Schi 257/1-4 and Schi
257/3-2), the EU (contracts SC1*-CT91-0642, CHRX-CT92-0051
and CHBG-CT94-0665) and PPARC (grant GR/K55738).

\end{document}